\documentclass{aa}
\usepackage{graphicx}
\usepackage{rotating}
\usepackage{afterpage}
\usepackage[sort&compress]{natbib}
\bibpunct{(}{)}{;}{a}{}{,}

\newcommand{\kms}{km\thinspace s$^{-1}$}

\newcommand{\Al}{{\it A$_\lambda$}}
\newcommand{\ks}{{\it K$_{\rm s}$}}
\newcommand{\Aks}{{\it A$_{\rm K_{\rm s}}$}}
\newcommand{\Ak}{{\it A$_{\rm K}$}}
\newcommand{\Av}{{\it A$_{\rm V}$}}
\newcommand{\sig}{$\sigma_{\rm A_{\rm K_{\rm s}}}$}

\usepackage{amsmath}
\def\new#1 {{\underline #1 }}
\def\cut#1 {\sout{#1} }

\begin{document}
   \title{86 GHz SiO maser survey of late-type stars in the Inner
   Galaxy \thanks{ This is paper no. 22 in a refereed journal based on
   data from the ISOGAL project.  Based on observations with ISO (an
   ESA project with instruments funded by ESA member states and with
   the participations of ISAS and NASA), with telescopes of the
   European Southern Observatory in La Silla, Chile, and with the IRAM
   30-m telescope on Pico Veleta, Spain. IRAM is supported by INSU/CNRS
   (France), MPG (Germany), and IGN (Spain). } 
\thanks{
Table 3 is only available in electronic form
at the CDS via anonymous ftp to cdsarc.u-strasbg.fr (130.79.128.5)
or via http://cdsweb.u-strasbg.fr/cgi-bin/qcat?J/A+A/}
 }

   \subtitle{III. Interstellar extinction and colours of the SiO targets} 
   \author{M.\ Messineo \inst{1,2} \and H.\ J.\ Habing
   \inst{2} \and K.\ M.\ Menten \inst{3} \and A.\ Omont \inst{4} \and
   L.\ O.\ Sjouwerman \inst{5} \and F.\ Bertoldi \inst{6}}

   \offprints{M. Messineo, \email{mmessine@eso.org}}

   \institute{European Southern Observatory, Karl Schwarzschild-Strasse 2, 
D-85748 Garching bei Munchen, Germany
         \and Leiden Observatory, P.O. Box 9513, 2300 RA Leiden, the Netherlands
         \and
             Max-Planck-Institut f\"ur Radioastronomie, Auf dem H\"ugel 69, D-53121 Bonn, Germany
         \and Institut d'Astrophysique de Paris, CNRS \& Universit\'e 
Paris 6, 98bis Bd Arago, F 75014 Paris, France
         \and
             National Radio Astronomy Observatory, P.O. Box 0, Socorro NM 87801, USA
 \and
Radioastronomisches Institut der Universit\"at Bonn, Auf dem H\"ugel 71,
              D-53121 Bonn, Germany
             }

   \date{Received xxxxxx xx, xxxx; accepted xxxxxx xx, xxxx}

\abstract{We have determined extinction corrections for a sample of
441 late-type stars in the inner Galaxy, which we previously searched
for SiO maser emission, using the 2MASS near-infrared photometry of
the surrounding stars.  From this, the near-infrared extinction law is
found to be approximated by a power law A$_\lambda \propto
\lambda^{\mathrm -1.9\pm0.1}$.  Near- and mid-infrared colour-colour
properties of known Mira stars are reviewed.  From the distribution of
the dereddened infrared colours of the SiO target stars we infer
mass-loss rates between $10^{-7}$ and $10^{-5}$ M$_\odot$
yr$^{-1}$.  } \maketitle
%
\section{Introduction}
We study interstellar extinction in the direction a sample of evolved
late-type stars in the inner Galaxy ($-4$\degr\ $< l < +30$\degr,
$|b|<1$\degr), which we previously searched for SiO maser emission
\citep[``SiO targets'' hereafter; ][ Paper\,I] {messineo02}.
\defcitealias{messineo02}{Paper\,I} Since SiO maser emission reveals
stellar line of sight velocities with an accuracy of a few \kms,
SiO maser stars are ideal for Galactic kinematics
studies. Furthermore, the combination of stellar kinematics and
physical properties of the stars, e.g.\ their intrinsic colours and
bolometric magnitudes, will enable a revised kinematic study of the
inner Galaxy to reveal which Galactic component and which epoch of
Galactic star formation the SiO stars are tracing.

A proper correction for interstellar extinction is of primary
importance for a photometric study of stellar populations in the inner
Galaxy, where extinction can be significant even at infrared
wavelengths. Interstellar extinction hampers an accurate determination
of the stellar intrinsic colours and bolometric magnitudes
\citep{thesis04}. This is especially critical in the central Bulge
region where interstellar extinction may exceed 30 visual magnitudes
and due to the current uncertainty in the near-infrared extinction law
(30\%) the uncertainty in the bolometric luminosities of evolved
late-type stars is at least 1 magnitude.

The available near- and mid-infrared photometry of the SiO targets
from the DENIS \citep{epchtein94}, 2MASS \citep{2massES}, ISOGAL
\citep{omont03,schuller03}, and MSX \citep{egan99,price01} surveys
were already presented by \citet[][Paper\,II]
{messineo03_2}. \defcitealias{messineo03_2}{Paper\,II} The sample
consists mainly of large-amplitude variable (LAV) Asymptotic Giant
Branch (AGB) stars \citepalias{messineo02,messineo03_2}.  The
estimates of interstellar extinction toward this class of objects are
complicated by the presence of a circumstellar envelope with variable
optical depth.  Therefore, in order to disentangle circumstellar and
interstellar extinction for each AGB star one also needs to study the
dust distribution along its line of sight.
 
In Sect.\ \ref{law} we discuss the uncertainty of the extinction law
at near- and mid-infrared wavelengths and the consequent uncertainty
of the stellar luminosities. In Sect.\ \ref{fieldextinction} we
describe the near-infrared colour-magnitude diagrams (CMDs) of field
stars (mainly giants) toward the inner Galaxy and use the latter to
derive the median extinction in the direction of each target. In
Sect.\ \ref{miras} we review the location of known Mira stars on the
CMDs and colour-colour diagrams. In Sects.\ \ref{siotargets} and
\ref{massloss} we use the median extinction from surrounding field
stars to deredden our SiO targets and discuss their colours and
mass-loss rates. The main conclusions are given in Sect.\
\ref{conclusion}.

\section{Interstellar extinction law}\label{law}
The composition and abundance of interstellar dust and its detailed
extinction properties remain unclear, thus limiting the accuracy of
stellar population studies in the inner Galaxy.  In the following we
discuss the near- and mid-infrared extinction law, in order to assess
the uncertainty in the extinction correction.

\begin{figure}[h!]
\begin{center}
\resizebox{\hsize}{!}{\includegraphics{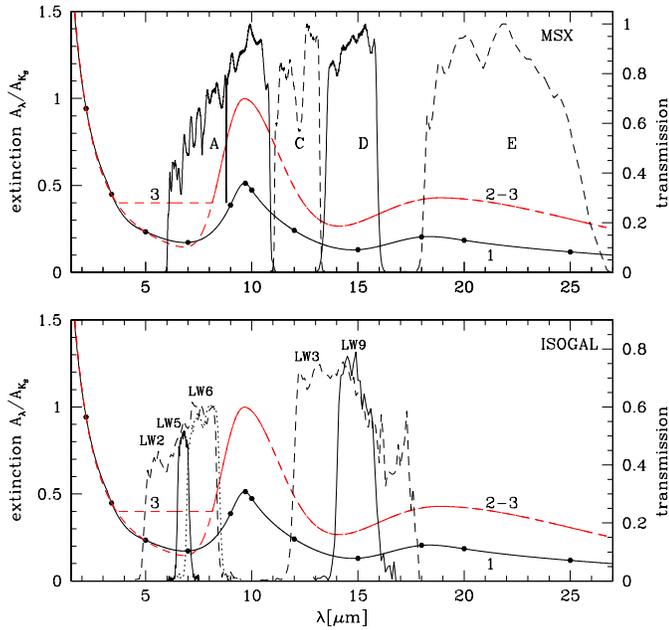}}
\end{center} 
\caption{\label{fig:filter.ps} Filter transmission curves and
extinction laws as function of wavelength. The continuous line (Curve
1) shows a fit to the values (dots) given by \citet{mathis90}; the
dashed curve shows the parametric expression given by
\citet{rosenthal00} plotted using a value of the silicate peak
$A_{9.7}/A_{2.12}=1.0$ (Curve 2). The latter is also shown without the
minimum around 4-8 $\mu$m (Curve 3), following \citet{lutz99}. {\bf In
the top panel} the transmission curves of the MSX $A, C, D$ and $E$
filters \citep{price01} are shown, and {\bf in the bottom panel} those
of the ISOCAM filters \citep{isocam03} used in the ISOGAL survey.}
\end{figure}

\subsection{Near-infrared interstellar extinction}\label{nearextinction}
\begin{table*}[th!]
\caption{\label{table:nearextinction} Near-infrared effective
extinction, $\langle$A$\rangle$/A$_{V} \propto \lambda^{-\alpha},$ for
various filters.  A different value of $R_{\rm V} = A_{\rm V}/E(B-V)$
does not affect {\rm A}$_{\rm K_{\rm s}}/{\rm E(H-K}_{\rm s})$ and
{\rm A}$_{\rm K_{\rm s}}/{\rm E(H-K}_{\rm s})$, but the slope of the
extinction curve does. Our findings favour a model with $\alpha=1.9$
(see Sect.\ \ref{fieldextinction}).}
\begin{tabular}{cccccclll}
\hline
\hline
{\rm A$_{\rm I}/{\rm A}_{\rm V}$}&{\rm A$_{\rm J}/{\rm A}_{\rm V}$}&{\rm A$_{\rm H}/{\rm A}_{\rm V}$}&{\rm A$_{\rm K_S}/{\rm A}_{\rm V}$}& {\rm A}$_{\rm K_S}/{\rm E(J-K}_{\rm s})$ &
{\rm A}$_{\rm K_{\rm s}}/{\rm E(H-K}_{\rm s})$&$\alpha$&{\rm R$_{\rm V}$} & Ref.\ \\
\hline
0.592&0.256&0.150&0.089& 0.533&1.459&1.85&    &\citet{glass99}\\
0.482&0.282&0.175&0.112& 0.659&1.778&1.61&3.09&\citet{rieke85}\\
0.606&0.287&0.182&0.118& 0.696&1.842&1.61&3.10&\citet{cardelli89} \\
0.563&0.259&0.164&0.106& 0.696&1.842&1.61&2.50$^{\mathrm{*}}$&''\\
0.606&0.267&0.158&0.096& 0.561&1.548&1.85&3.10&\citet{cardelli89}$^{\mathrm{+}}$\\
0.563&0.240&0.142&0.086& 0.561&1.548&1.85&2.50&''\\
0.606&0.263&0.153&0.092& 0.537&1.496&1.90&3.10&''\\
0.563&0.237&0.138&0.083& 0.537&1.496&1.90&2.50&''\\
0.606&0.255&0.144&0.084& 0.493&1.401&2.00&3.10&''\\
0.563&0.229&0.130&0.076& 0.494&1.400&2.00&2.50&''\\
0.606&0.238&0.127&0.070& 0.420&1.236&2.20&3.10&''\\
0.563&0.213&0.114&0.063& 0.420&1.236&2.20&2.50&''\\ \hline
\end{tabular}
\begin{list}{}{}
\item[$^{\mathrm{*}}$] recent determination toward the Bulge 
\citep[e.g.][]{udalski03}.
\item[$^{\mathrm{+}}$] parametric expression modified to 
extrapolate to $\lambda >0.9~ \mu$m with $\lambda^{-\alpha}$.
\end{list}
\end{table*}

Interstellar extinction at near-infrared wavelengths (1-5 $\mu$m) is
dominated by graphite grains.  Near-infrared photometric studies have
shown that the wavelength-dependence of the extinction may be
expressed by a power law {\it \Al} $\propto \lambda^{-\alpha}$ where
$\alpha$ was found to range between 1.6 \citep{rieke85} and 1.9
\citep{glass99, landini84,vandehulst46}.

One can estimate the near-infrared extinction by measuring the
near-infrared reddening of stars of known colour.  When using
broad-band photometric measurements, one needs to properly account for
the bandpass, stellar spectral shape, and the wavelength-dependence of
the extinction.  As a function of the \ks\ band extinction we
therefore computed a grid of ``effective extinction'' values for the
DENIS $I$ and 2MASS $J,~H$, and $K_{\rm s}$ bandpass. This effective
extinction was computed by reddening an M0 III stellar spectrum
\citep{fluks94} with a power law extinction curve and integrating over
the respective filter transmission curves. When we convolve the filter
response with a stellar sub-type spectrum different from the M0 III,
only the effective I-band extinction values change significantly,
e.g.\ decreasing by 3\% for a M7 III spectrum \citep[see
also][]{vanloon03}.

The \ks-band extinction \Aks\ can then be expressed as
$$A_{\rm K_{\rm s}} = C_{JK} \times E(J-K_{\rm s}),$$
$$A_{\rm K_{\rm s}} = C_{HK} \times E(H-K_{\rm s}),$$ where
$E(J-K_{\rm s})$ and $E(H-K_{\rm s})$ are the reddening in the
$J-K_{\rm s}$ and $H-K_{\rm s}$ colour, respectively, and the $C$ are
constants.  These relations are independent of visual extinction and
of the coefficient of selective extinction, $R_{\rm V} = A_{\rm
V}/E(B-V)$, but they depend on the slope of the near-infrared
extinction power law (see Table \ref{table:nearextinction}).  However,
to provide the reader with the familiar ratio between near-infrared
effective extinction and visual extinction, we also used the commonly
adopted extinction law of \citet{cardelli89}.  Such ratios may be
useful in low-extinction Bulge windows, where visual data are also
available.   The analytic expression given by Cardelli et al.,
which depends only on the parameter $R_{\rm V}$, is based on
multi-wavelength stellar colour excess measurements from the violet to
0.9$\mu$m, and extrapolates to the near-infrared using the power law of
\citet{rieke85}. We extrapolated Cardelli's extinction law to
near-infrared wavelengths using a set of different power laws (Table
\ref{table:nearextinction}).

Uncertainty in the slope of the extinction law produces an uncertainty
in the estimates of the near-infrared extinction of typically 30\% in
magnitude (see Table \ref{table:nearextinction}).  For a \ks\ band
extinction of \Aks$=3$ mag the uncertainty may be up to 0.9 mag, which
translates into an uncertainty in the stellar bolometric magnitudes of
the same magnitude.

In Sect.\ \ref{fieldextinction} we show that a power law index
$\alpha=1.6$ is inconsistent with the observed near-IR colours of
field giant stars toward the inner Galaxy, and that the most likely
value of $\alpha$ is $1.9\pm 0.1$.

\subsection{Mid-infrared interstellar extinction}
The mid-infrared extinction (5-25 $\mu$m) is characterised by the 9.7
and 18 $\mu$m silicate features. The strength and profile of these
features are uncertain and appear to vary from one line of sight to
another.  A standard graphite-silicate mix predicts a minimum in
$A_{\lambda}/A_{2.12}$ at 7$\mu$m, which has however been not observed
to be very pronounced toward the Galactic Centre
\citep{lutz96,lutz99}.

\begin{table*}
\caption{\label{table:extinction} Effective extinction,
$\langle$A$\rangle$/A$_{K_S}$, using M-giant spectra \citep{fluks94}
for different bands defined by the ISOCAM and MSX filters (see Fig.\
\ref{fig:filter.ps}). $A_{K_S}/A_{2.12}=0.97$. }
\begin{tabular}{lrrcccc}
\hline
\hline
{\rm Filter}& $\lambda_{\rm ref}$&$\Delta \lambda$ & {\rm Curve 1 (Mathis)} &{\rm Curve 2 }&{\rm Curve 3 (Lutz)}\\
            &                    &                 & {\rm  ($A_{9.7}/A_{2.12}=0.54$)}   &{\rm  ($A_{9.7}/A_{2.12}=1.00$)}&{\rm ($A_{9.7}/A_{2.12}=1.00$ \& no minimum)}\\
{\rm }&{\rm $\mu$m }& {\rm  $\mu$m}&{\rm $\langle$A$\rangle$/A$_{K_S}$ }& {\rm $\langle$A$\rangle$/A$_{K_S}$ }&{\rm $\langle$A$\rangle$/A$_{K_S}$}\\
\hline
{\rm LW}2& 6.7&3.5& 0.21&  0.21&0.41\\
{\rm LW}5& 6.8&0.5& 0.18&  0.15&0.41\\
{\rm LW}6& 7.7&1.5& 0.21&  0.26&0.43\\
{\rm LW}3&14.3&6.0& 0.18&  0.34&0.34\\
{\rm LW}9&14.9&2.0& 0.14&  0.29&0.29\\
{\rm A}  &8.28&4.0& 0.26&  0.38&0.55\\
{\rm C}  &12.1&2.1& 0.25&  0.49&0.49\\
{\rm D}  &14.6&2.4& 0.14&  0.29&0.29\\
{\rm E}  &21.3&6.9& 0.17&  0.41&0.41\\
\hline
\end{tabular}
\end{table*}

The mid-infrared extinction curve of \citet{mathis90} is commonly used
with its combination of a power law and the astronomical silicate
profile of \citet{draine84} with $A_{9.7}/A_{2.2} \simeq 0.54$ -- a
value found in the diffuse interstellar medium toward Wolf-Rayet stars
\citep[e.g.][]{mathis98}. However, from observations of hydrogen
recombination lines, \citet{lutz99} concluded that $A_{9.7}/A_{2.2}
\simeq 1.0$ in the direction of the Galactic centre; and analysing the
observed H$_2$ level populations toward Orion OMC-1,
\citet{rosenthal00} derived $A_{9.7}/A_{2.12} = 1.35$.

By using a parametric mid-infrared extinction curve by
\citet{rosenthal00}, we computed two curves with
$A_{9.7}/A_{2.2}=$1.0, one in combination with the minimum at 4-8
$\mu$m (Curve 2) and one without it as suggested by \citet{lutz99}
(Curve 3) (see Fig.\ \ref{fig:filter.ps}). Using these extinction laws
we computed the effective extinctions $\langle A \rangle$/\Aks\ for
the M-type synthetic spectra \citep{fluks94} in the various filters
used by the ISOGAL and MSX surveys (Table \ref{table:extinction}).
The ratios $\langle A \rangle$/\Aks\ are not sensitive to the stellar
sub-type used.  An increase in the ratio $A_{9.7}/A_{2.2}$ from 0.54
to 1.0 results in an increase between 0.15 and 0.20$\times$\Aks\ of
the average attenuation in the $LW3,~LW9,~C,~D$, and $E$ spectral
bands.  The spectral bands of the $LW2$ and $LW5$ filters are not very
sensitive to the intensity of the silicate feature, but to the minimum
of the extinction curve in the 4-8 $\mu$m region. Although $\langle A
\rangle$/\Aks\ varies with \Aks, these variations are small compared
with those arising from different choices of the mid-infrared
extinction law.

In the most obscured regions (\Aks$=3$) uncertainties for the
ISOGAL and MSX filters range from 0.45 mag ($LW3,LW9,D$) to 0.85 mag
($A$).  Because their energy is emitted mostly at near-infrared
wavelengths, this has a negligible effect (0.1 mag in average) on the
calculated $M_{\mathrm {bol}}$ of the SiO targets and, therefore,
also on the mass-loss rate estimates (see Sect.\
\ref{massloss}). In the following we use the Lutz law (Curve 3), since
it ensures  consistency between mid-infrared and near-infrared
stellar colours \citep{jiang03}.

\section{Interstellar extinction of field stars derived from 
near-infrared colour-magnitude diagrams}
\label{fieldextinction}

\begin{figure*}[th!]
\resizebox{0.5\hsize}{!}{\includegraphics{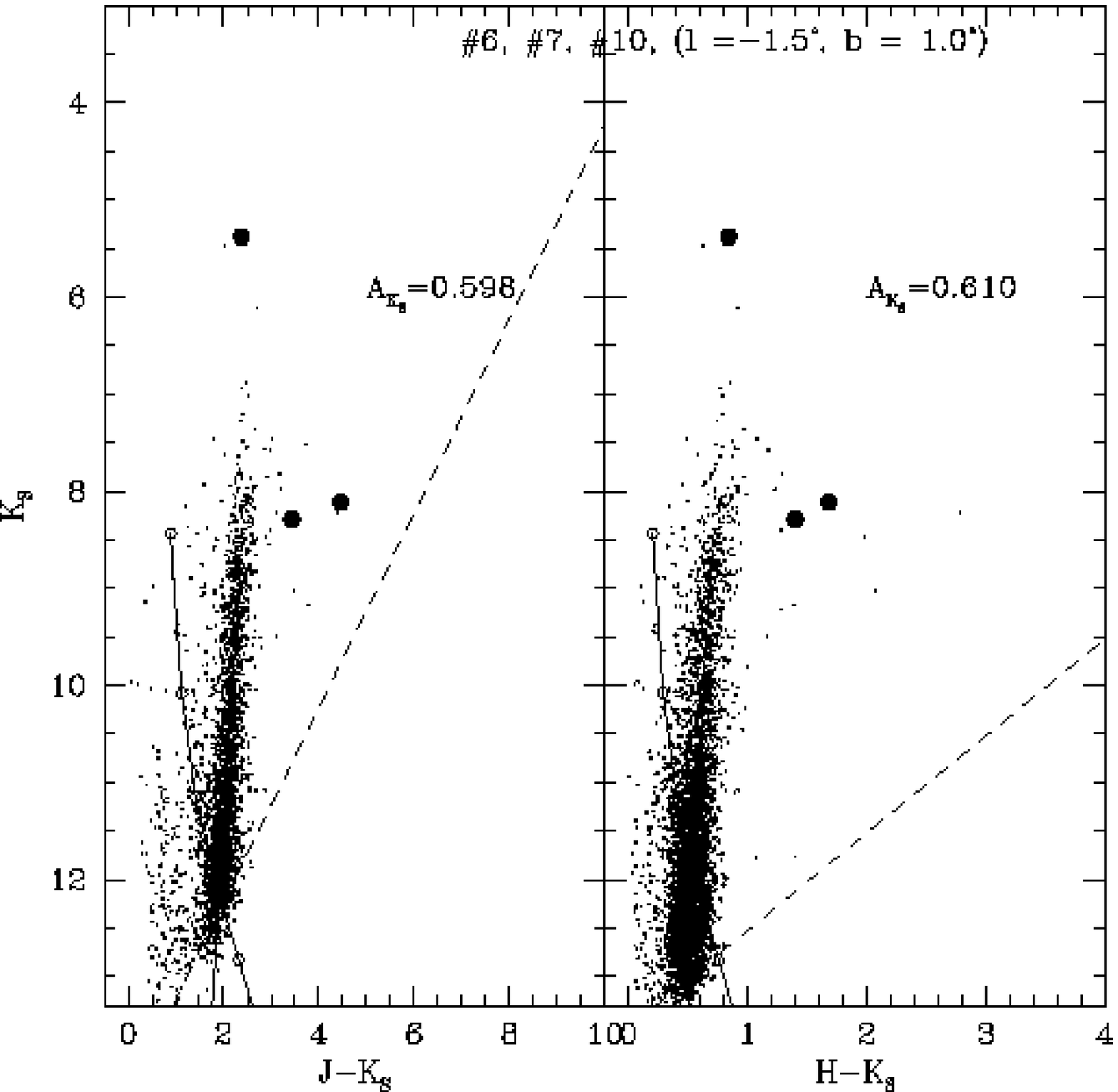}}
\resizebox{0.5\hsize}{!}{\includegraphics{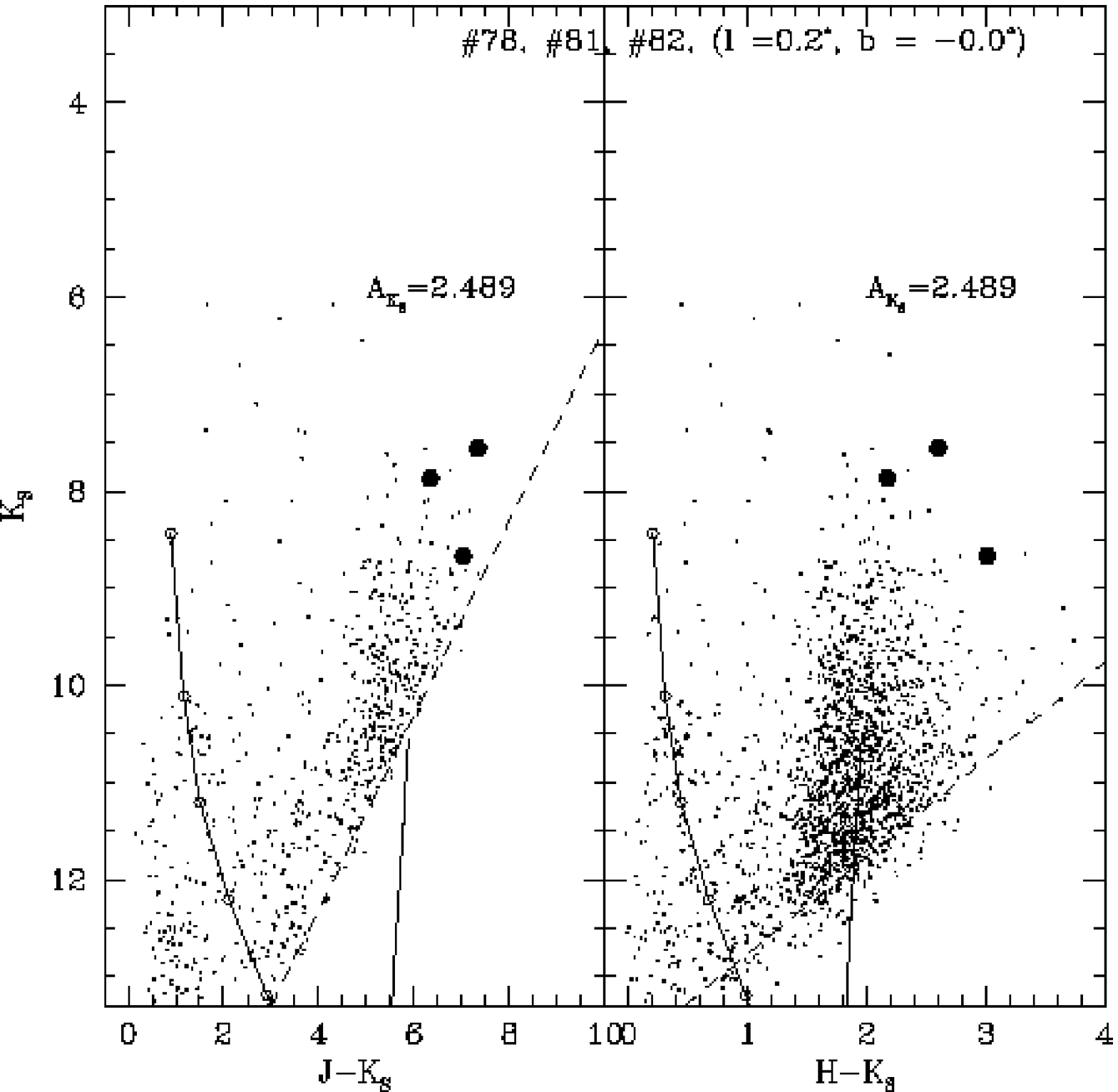}}
\resizebox{0.5\hsize}{!}{\includegraphics{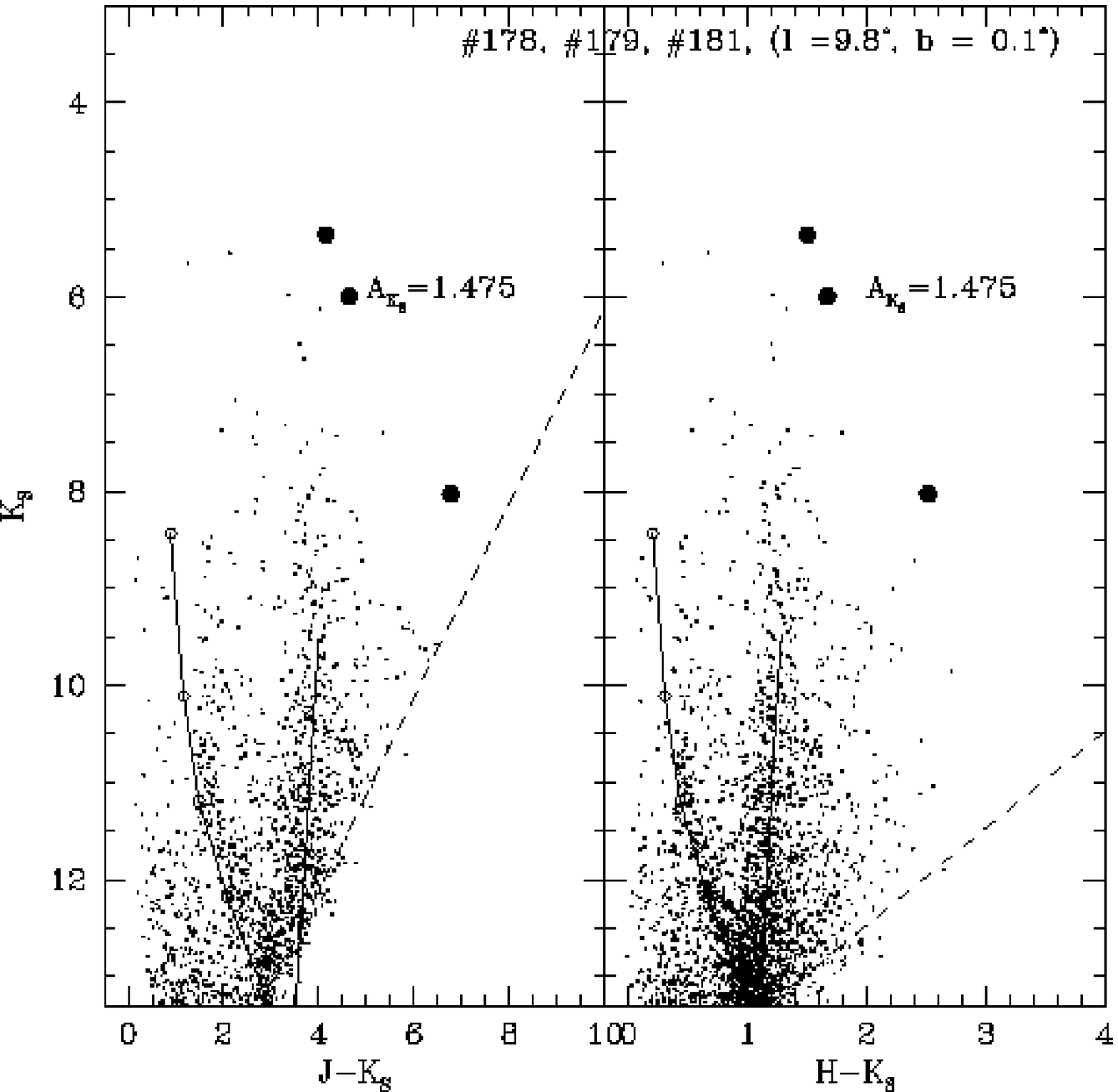}}
\resizebox{0.5\hsize}{!}{\includegraphics{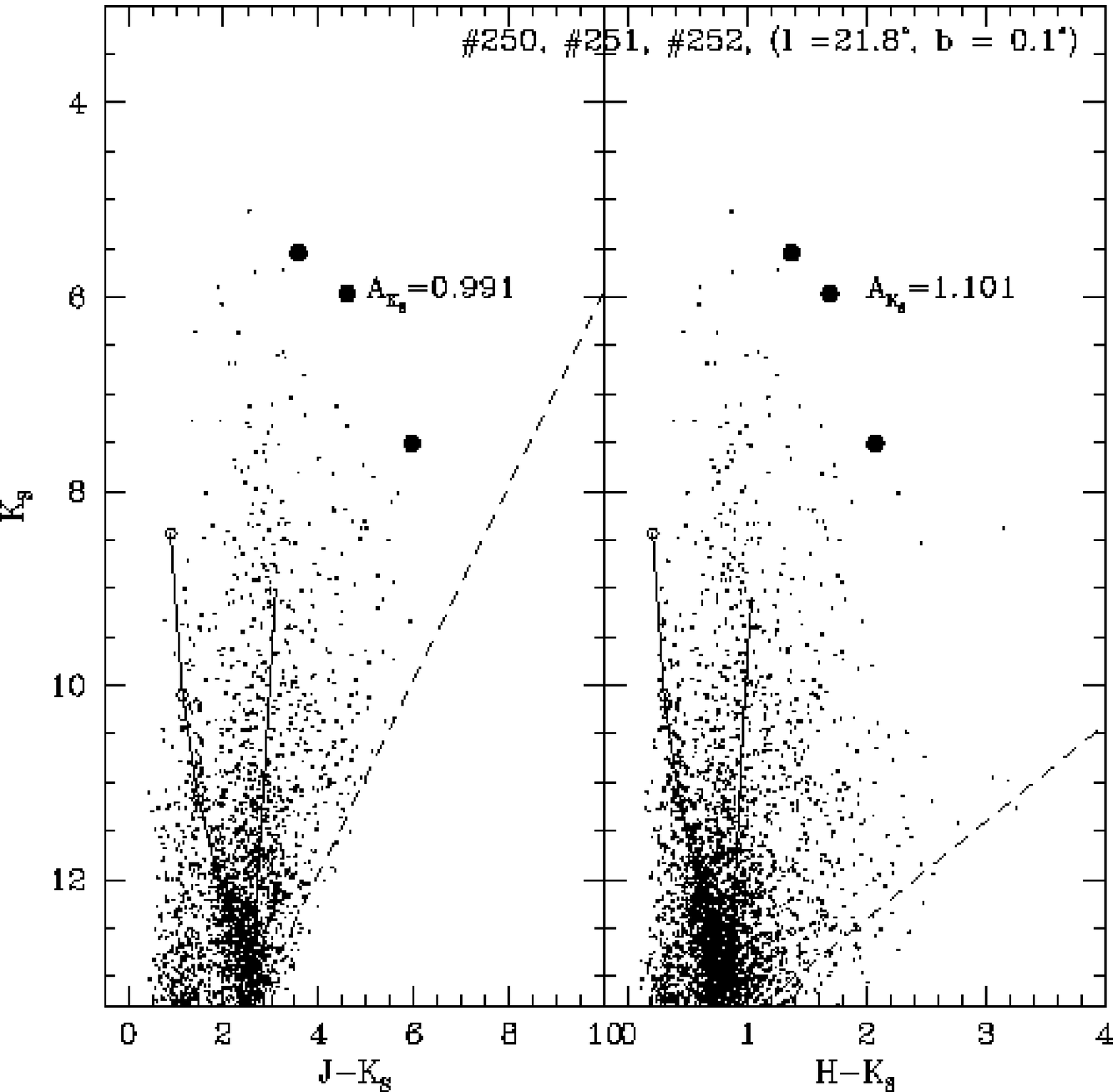}}
\caption{\label{fig:panels} Colour-magnitude diagrams of 2MASS
datapoints (small dots) of good quality (0 $<$ flag-red $\leq$ 3)
located within 4\arcmin\ from the position of the SiO target (big
dot). Three fields at equal median extinction are combined in each
panel. The right-hand continuous line indicates the locus of the
reference RGB curve (see Sect.\ \ref{rgb}), adopting a distance of 8
kpc and reddening it with the median extinction of field stars
(\Aks). The left-hand continuous curve shows the tracels of clump stars
for increasing distance and extinction along a given line of sight
(see Sect.\ \ref{outside}), obtained using the extinction model by
\citet{drimmel03} and the absolute magnitudes from
\citet{wainscoat92}. Dashed lines indicate the diagonal cut-off due to
the detection limits in $J$ and $H$. Circles on the clump trace mark a
distance from 1 to 5 kpc with a step of 1 kpc downward.}
\end{figure*}

Most of the sources detected by DENIS and 2MASS toward the inner
Galaxy are red giants and AGB stars. Because the intrinsic colours of
giants are well known and steadily become redder with increasing
luminosity, one can study the near-infrared CMDs to estimate the
average extinction toward a given line of sight for a population of
such stars. This approach has been used to map the extinction in the
central region of the Galaxy ($|l|<10$\degr)
\citep{schultheis99,dutra03}.

Assuming the SiO target stars to be spatially mixed with the red giant
stars, we can similarly estimate extinction, \Aks, toward the 441
SiO targets.  In both the (\ks, $J-$\ks) and (\ks, $H-$\ks) diagrams
the colour and magnitude of field stars within a 2-4\arcmin\ radius
from each SiO target were shifted to the reference red giant branch,
RGB, (see Sect.\ \ref{rgb}) along the reddening vector. Selection
effects due to magnitude limits were taken into account (Sect.\
\ref{planes}). Median extinction of the field was thereby determined
in both the (\ks, $J-$\ks) and (\ks, $H-$\ks) planes. To exclude
foreground stars, an iterative 2$\sigma$ clipping was applied to the
extinction distribution \citep{dutra03}.

Nearly 90\% of the SiO target stars appear redder than their
neighbouring field stars (Fig.\ \ref{fig:panels}), which implies that
the SiO stars are intrinsically obscured, provided their spatial
distribution is the same as that of RGB stars.

\subsection{Reference red giant branch}\label{rgb}
Our method for measuring the extinction needs reference to an RGB: we
adopted that of 47 Tuc, brought to 8 kpc by adopting a distance
modulus of 13.32 \citep{ferraro99}. This choice is supported by a
recent new near-infrared colour-magnitude analysis of the RGB in
globular clusters by \citet{valenti04}, who obtained $(J,H,K)$
photometry taken at the ESO 2.2 m telescope with a spatial resolution
of 0.3 and 0.5\arcsec\ pixel$^{-1}$, half that of 2MASS. They
showed that within 0.05 mag the RGB of 47 Tuc and the metal-rich Bulge
globular clusters (such as NGC 6528, NGC 6553, NGC 6540) have
identical colours.
 
In the 2MASS $K_{\rm S0},(J-K_{\rm s})_0$ CMD, the upper part
(\ks$<12$ mag) of the 47 Tuc RGB is well represented by a linear fit
\begin{equation}
({\rm J-K_{\rm s}})_0 = 2.19(\pm0.02)-0.125(\pm0.002){\rm K_{S0}};
\end{equation}
in the $(H-K_{\rm s})_0$ vs. $K_{s0}$ diagram a second-order
  polynomial fits well the RGB (\ks$_0 < 12$ mag):
\begin{equation}
({\rm H-K_{\rm s}})_0 = 1.73(\pm0.22)-0.268(\pm0.05)\times{\rm K_{S0}}
\end{equation} 
\vspace{-0.4 cm}
$$+0.011(\pm0.002)\times{\rm K_{S0}^2}.$$ 

The $K_{\rm s},(H-K_{\rm s})$ plane is less sensitive to extinction
than the $K_{\rm s},(J-K_{\rm s})$ plane: while a colour change of 0.1
in $(J-K_{\rm s})_0$ implies a change of 0.05 in \Aks\ (0.6 in \Av), a
change of 0.1 mag in $(H-K_{\rm s})_0$ implies a change in \Aks $\sim
0.15$. A shift in distance modulus of the reference RGB of $\pm 2$ mag
results in a small change in the extinction of \Aks$\mp 0.15$ mag.

\begin{table}[ht!] 
\begin{center}
\caption{\label{table:final} Extinction values.  The identification
number (ID) of the SiO target, as in Table 2 and 3 of
\citetalias{messineo02}, is followed by the field extinction \Aks, by
the corresponding dispersion of individual extinctions of field stars
and by the total (circumstellar plus interstellar) extinction in
\ks-band toward the target star (tot). The flag (Fg) is unity when the
SiO target is classified as a ``foreground object''.}  {\footnotesize
\begin{tabular}{@{\extracolsep{-.04in}}rrrrr}
\hline 
\hline 
{\rm ID} & {\Aks} &{\sig} &{\rm tot}& {\rm Fg}\\
{\rm   } & {\rm mag} &{\rm mag}&{\rm mag} &    \\
\hline 

  1& 0.96& 0.18& 1.63&   \\ 
  2& 1.23& 0.31& 2.81&   \\ 
  3& 1.90& 0.33& 1.79&   \\ 
  4& 2.16& 0.52& 3.00&   \\ 
  5& 1.51& 0.25& 1.75&   \\ 
  6& 0.59& 0.09& 1.16&   \\ 
  7& 0.57& 0.07& 0.51&   \\ 
  8& 1.98& 0.41& 1.53&1  \\ 
  9& 2.14& 0.32& 2.59&   \\ 
 10& 0.61& 0.11& 1.48&   \\ 
\hline 
\end{tabular}
\begin{list}{}{}
\item[$^{\mathrm{*}}$] The full table is available in electronic form
at the CDS via anonymous ftp to cdsarc.u-strasbg.fr (130.79.128.5) or
via  ${\rm http}://{\rm cdsweb.u-strasbg.fr/cgi}-{\rm
bin/qcat?J/A+A/(vol)/(page)}$.
\end{list}
}
\end{center}
\end{table} 

\subsection{Determination of extinction value and extinction law 
in the $J$, $H$, $K_{\rm s}$ CMD}
\label{planes}
To properly estimate the interstellar extinction along a given line of
sight one should consider only the region of the CMD where the upper
RGB is well defined, i.e., not affected by large photometric errors or
by the diagonal cut-off from the 2MASS detection limits (see Fig.\
\ref{fig:panels}) that would bias the calculation of the median
extinction toward a lower value \citep[see
also][]{dutra03,cotera00,figer04}.

In the inner Galaxy, the photometric error is typically smaller than
0.04 for stars with $K_{\rm s}<12$ mag.  To quantify the
incompleteness due to the diagonal cut-off with zero extinction, our
average 2MASS detection limits of $J=16.0$ and $H=14.0$ correspond to
a RGB \ks\ magnitude of 15.2 and 13.0 mag, respectively, at a distance
of 8 kpc. Accounting for scatter in the observed colours of $\pm 0.5$
mag, the RGB would be well sampled to $J=15.5$ and $H=13.5$ mag,
corresponding to \ks$< 14.6$ and 12.6, respectively.  With a \ks\
extinction of 3 mag ($\sim 5.0$ mag in $H$, $\sim 8.6$ mag in $J$),
which is a typical value toward the Galactic centre, these RGB
completeness limits would rise to $K_{\rm s}=7.1$ and 10.8 mag in the
$(K_{\rm s},J-K_{\rm s})$ and $(K_{\rm s},H-K_{\rm S})$ planes,
respectively. Therefore only the $H$ band can provide a sufficient
number of red giant stars to match the reference RGB. With the 2MASS
data $(K_{\rm s},J-K_{\rm s})$ CMDs are useful for extinction
determinations only to a \ks\ extinction of about 1.6 mag.

For fields with \Aks $<1.6$ mag Table \ref{table:final} lists the
extinction values determined from the (\ks,$J-$\ks) plane, at larger
extinction values those from the (\ks,$H-$\ks) plane.

\begin{figure}[h!]
\begin{center}
\resizebox{0.8\hsize}{!}{\includegraphics{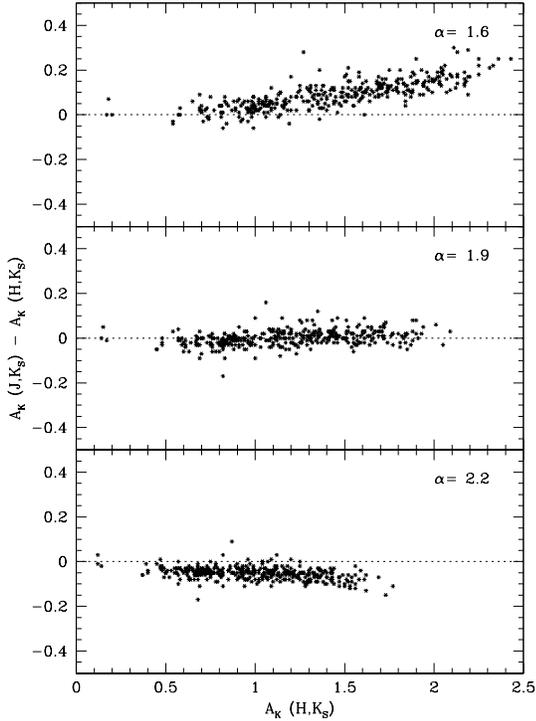}}
\caption{\label{extall.ps} Comparison between the median extinction
obtained from the (\ks,$J-$\ks) and
 (\ks,$H-$\ks) planes.  Only sources detected in
$J,H,$\ks\ above the \ks\ completeness limits are used.  In each panel
a different value of the spectral index of the extinction power law,
$\alpha$, is adopted. }
\end{center}
\end{figure}

By using both the (\ks,$J-$\ks) and (\ks,$H-$\ks) planes one can in
theory fit the observed field star colours to the RGB and fit both the
absolute average $K_{\rm s}$ band extinction and the index, $\alpha$,
of the extinction power law.  For this we selected field stars which
were detected in $J$, $H$, and \ks and which are brighter than the
\ks\ completeness limits for the RGB at the extinction of each
field. For each field we determined the median of the extinctions of
the field stars in both the $(K_{\rm S},J-K_{\rm s})$ and in the
$(K_{\rm s},H-K_{\rm s})$ CMDs. The power law index $\alpha$ was
chosen such that the extinctions agreed in the two planes.  For
different values of $\alpha$, Fig. \ref{extall.ps} shows the
differences in median extinction values \Aks($J$,\ks)$-$\Aks($H$,\ks)
plotted against \Aks.  Only for $\alpha=1.9\pm0.1$ are the two
extinction estimates found to be consistent.

In the following we use $\alpha = 1.9$, which is consistent with the
work of \citet{glass99} and \citet{landini84}, and with the historical
Curve 15 of \citet{vandehulst46}.

\subsection{Outside the Bulge}\label{outside}
K2 giant stars are the dominant population of late-type stars seen
along the galactic disk \citep[e.g.][]{drimmel03,lopez03}. They
correspond to red clump stars in metal-rich globular clusters such as
47 Tuc.  The location of clump stars on the CMD is marked in Fig.\
\ref{fig:panels}.

Toward the Bulge, the Bulge RGB population is dominant; therefore, the
median interstellar extinction is not affected by possible foreground
clump stars.  This is not the case in the disk ($l = \sim10$\degr),
where one must ignore the foreground clump stars before fitting the
RGB. As likely clump stars we identify those within 0.3 mag from the
$J-$\ks\ colour of the clump trace, and as giants those stars redder
than the clump stars \citep[e.g.][]{lopez03}.

\subsection{Dispersion  of the extinctions along a line of sight}
Along with the median extinction of the field stars, \Aks, we also
determine the standard deviation of the distribution of the individual
extinction estimates, \sig. The patchy nature of the extinction is
apparent even within the $2-4\arcmin $ radius sampling area. This
patchiness integrated over a longer path generates larger \sig\ with
increasing extinction for Bulge lines of sight.  The 1\sig\
uncertainty in the field extinction varies from $\sim0.2$ mag when
\Aks$ = 0.6$ mag up to $\sim0.7$ mag in the regions with the largest
extinction (\Aks$>2.0$).  In fields at longitudes longer than 10\degr,
a larger \sig\ is found than in Bulge fields of similar median
extinction. This is probably due to the presence of several Galactic
components, e.g.\ the disk, arms, bar, and molecular ring.

\section{Near-infrared properties of known Mira stars}
\label{miras}
Presently the pulsation periods and amplitudes of our SiO targets are
unknown, although most of them must be LAVs
\citepalias{messineo03_2}. This is suggested by photometric variations
seen in measurements taken at different times (DENIS, 2MASS and MSX)
by the strong 15 $\mu$m emission \citepalias{messineo03_2} and by the
SiO maser emission \citepalias{messineo02}.

Although they are 20 times less numerous than semiregular AGB stars
(SRs) \citep{alard01}, Mira stars are among the best studied pulsating
variable stars. They are regular long period variables (LPV) with
visual amplitudes larger than 2.5 mag and $K$ band amplitudes
exceeding $\sim 0.3$ mag (LAV stars). Since large amplitudes tend to
be associated with the most regular light curves \citep{cioni03},
amplitude is the best choice for classifying Mira stars.

To analyse the colours of our SiO targets, in particular to check the
quality of the extinction corrections, it is useful to have comparison
samples of well studied large amplitude LPV stars with a wide range of
intrinsic colours, at low interstellar extinction and at different
Galactic positions.  Therefore, we examined: two samples of Mira stars
in the solar vicinity \citep{olivier01,whitelock-hyp}; one sample
toward the Galactic Cap taken from \citet{whitelock94}; 18 Mira stars
detected by IRAS \citep[][]{glass95} in the Sgr--I field; 104 IRAS
Mira stars in the outer Bulge \citep{whitelock91}; and a sample of LPV
in the Large Magellanic Cloud \citep{whitelock03}.  All these stars
have IRAS 12 $\mu$m magnitude [12] and mean $J,H,K$ magnitudes in the
SAAO system \citep{carter90}.

Stellar fluxes are given already corrected for reddening only in the
work of \citet{olivier01}.  For the LPV stars analysed by
\citet{whitelock03, whitelock-hyp,whitelock94} the effects of
interstellar extinction are negligible because these stars are nearby
or outside of the Galactic plane, so we did not correct these for
extinction. We dereddened the Baade Sgr--I window data \citep{glass95}
adopting our favourite extinction curve ($\alpha = 1.9$) and \Aks\ $=
0.15$ mag, consistently with the value adopted in
\citet[][]{glass95}. We corrected the magnitudes of the outer Bulge
Miras \citep{whitelock91} for reddening adopting values of \Aks\
derived from their surrounding stars (see next Sect.).  Next we
analyse the location of these well-known Mira stars in the
near-infrared CMDs and colour-colour diagrams.

\subsection{Colour-magnitude diagram of outer Bulge Mira
stars and  surrounding field stars}
\begin{figure}[th!]
\begin{center}
\resizebox{0.8\hsize}{!}{\includegraphics{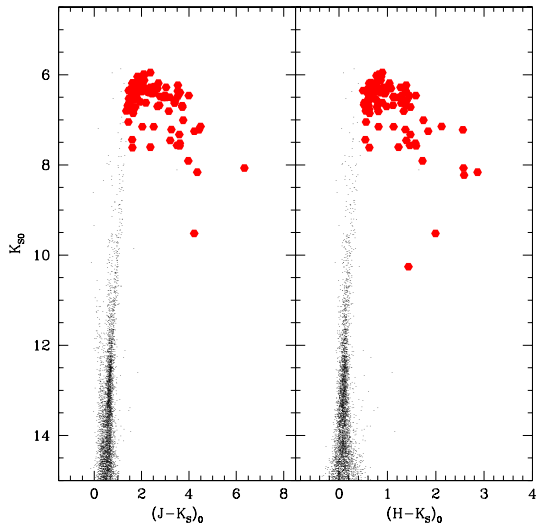}}
\caption{\label{fig:whitelock91.ps} Dereddened colour-magnitude
diagrams. Big dots represent the outer Bulge Mira stars found by
\citet{whitelock91}. The magnitudes plotted are mean magnitudes at the
equal distance of 8 kpc, adopting the distances of
\citet{whitelock91}. Small dots represent the point sources detected
by 2MASS within 1\arcmin\ from each Mira star.}
\end{center}
\end{figure}
The outer Bulge LPV stars studied by \citet[][]{whitelock91} are
particularly interesting for a comparison with our SiO targets because
they were selected on the basis of their IRAS fluxes and colours with
criteria similar to those used to select our MSX targets
\citepalias{messineo02,messineo03_2}. Since their main periods range
from 170 to 722 days and their $K$ amplitudes from 0.4 to 2.7 mag,
they are classical Mira stars. Their distribution of distance moduli,
estimated from the period-luminosity relation \citep{whitelock91},
peaks at 14.7 mag with a $\sigma = \sim0.5$ mag. Since they are at
latitudes between 6 and 7\degr, they are in regions of low
interstellar extinction. All this makes them ideal for a comparison
with our SiO targets, the study of which is complicated by the large
interstellar extinction at their low latitudes.

An RGB is clearly apparent in Fig.\ \ref{fig:whitelock91.ps}, which
shows the 2MASS point sources within 1\arcmin\ of each Mira star.  The
median extinction toward each field ranges from \Aks= 0.01 to 0.30 mag
with a typical dispersion of 0.01-0.08 mag.  On the CMDs the Mira
stars appear mostly brighter than the RGB tip of the field stars
\citep[$K = 8.2$ mag at a distance of 8 kpc, see][]{frogel87}.  Due to
the presence of a circumstellar envelope, Mira stars have red colours
(up to $(H-K_{\rm s})_0 = 3$ mag) and lie on the red-side of the giant
sequence. It is therefore not possible to derive the interstellar
extinction toward Mira stars from their colours relative to the RGB.

There is no reason to assume the Mira stars are spatially distributed
differently from other giant stars, so that the extinction toward the
surrounding field stars may serve as an approximation of that of the
respective Mira star. One worry in this assumption is not knowing the
actual distribution of extinction along the line of sight.

\begin{figure}[h!]
\begin{center}
\resizebox{0.8\hsize}{!}{\includegraphics{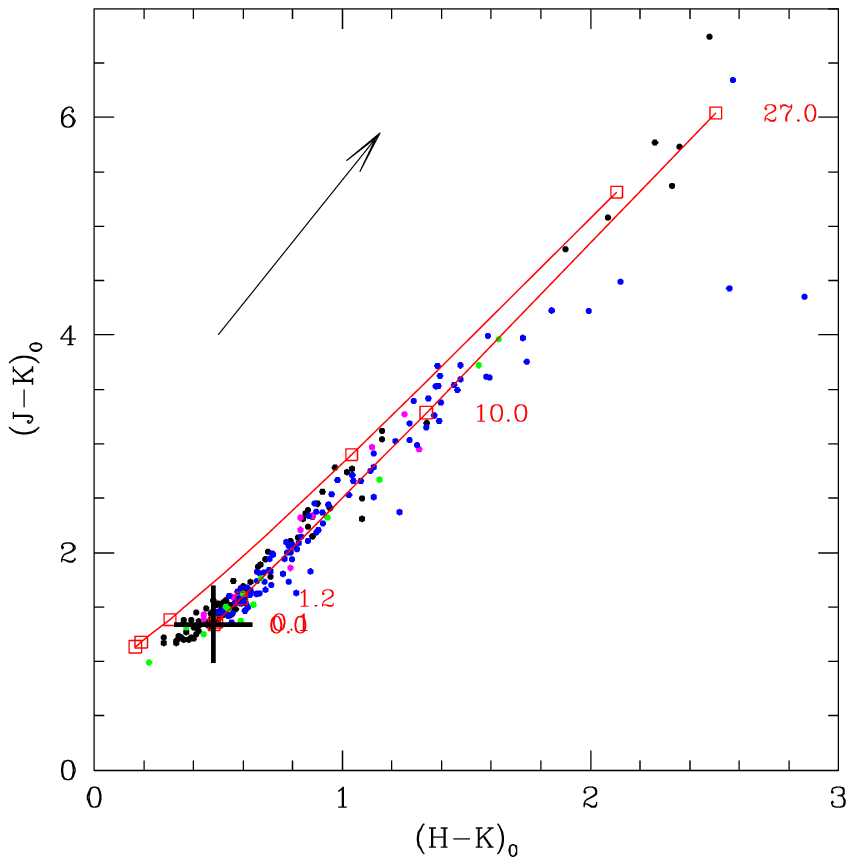}}
\caption{\label{fig:jhkLPV.ps} Dereddened colours of known  Mira
stars \citep{olivier01,whitelock-hyp,whitelock94,glass95,whitelock91}.
Near-infrared mean magnitudes are used.  The two curves represent M3
(upper) and  M10 (lower) type stars with increasing mass-loss rates
(indicated by squares and labels $\times 10^{-6}$ M$_\odot$/year), as
modelled by \citet{groenewegen93}.  The cross indicates the position
of an M10 star without mass-loss ($(H-K)_0 = 0.48$ and
$(J-K)_0 = 1.34$). The arrow shows the reddening vector for \Ak$ = 1$
mag.}
\end{center}
\end{figure}

\subsection{Colour-colour diagram of Mira stars}
\begin{figure}[h!]
\begin{center}
\resizebox{0.8\hsize}{!}{\includegraphics{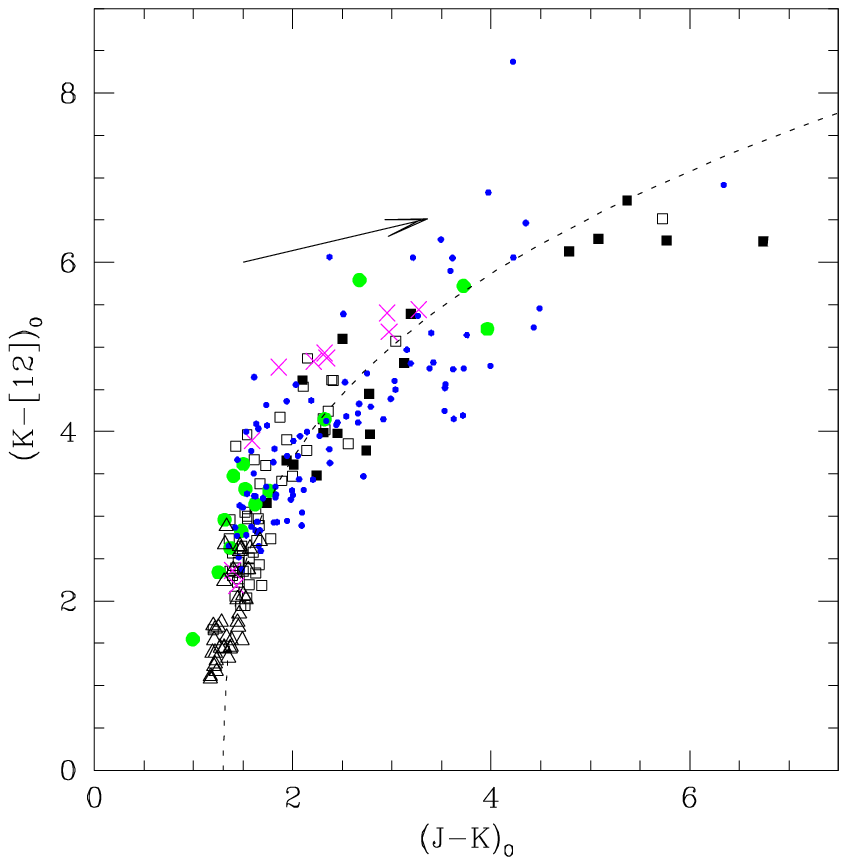}}
\caption{\label{fig:jk12LPV.ps} Dereddened colours of
         infrared-monitored Mira stars (based on the IRAS 12 $\mu$m
         and near-infrared mean magnitudes): in the solar vicinity
         (filled squares) \citep{olivier01}; detected by Hipparcos
         (open triangles) \citep{whitelock-hyp}; toward the South
         Galactic Cap (open squares) \citep{whitelock94}; in the Baade
         Sgr--I window (big dots) \citep{glass95}; in the outer Bulge
         (small dots) \citep{whitelock91}; in the Large Magellanic
         Cloud (crosses) \citep{whitelock94}.  The dotted line is the
         best fit to an IRAS sample of oxygen-rich AGB stars
         \citep{vanloon97}.  The arrow shows the reddening vector for
         \Ak$ = 1$ mag.  }
\end{center}
\end{figure}
\begin{figure}[h!]
\begin{center}
\resizebox{0.8\hsize}{!}{\includegraphics{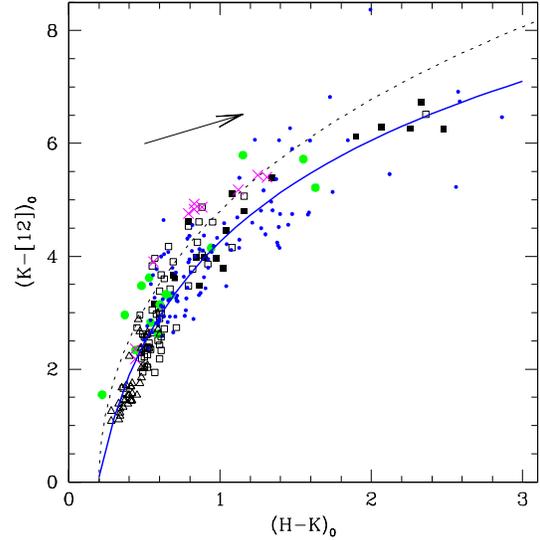}}
\caption{\label{fig:hk12LPV.ps} Dereddened colours of
         infrared-monitored Mira stars, based on the IRAS 12 $\mu$m
         and near-infrared mean magnitudes. Symbols are the same as in
         Fig.\ \ref{fig:jk12LPV.ps}.  The dotted line is the best fit
         to an IRAS sample of oxygen-rich AGB stars \citep{vanloon98}. 
         The continuous curve is our best fit for
         Galactic Mira stars. The arrow shows the reddening vector for
         \Ak$ = 1$ mag.}
\end{center}
\end{figure}
For Mira stars with low mass-loss rate ($<10^{-7}$ M$_\odot$
yr$^{-1}$), the $(J-K)_{0}$ colours range between 1.2 and 1.6
\citep{whitelock-hyp}.  Dust-enshrouded IRAS AGB stars with mass-loss
rates of $10^{-6} - 10^{-4}$ M$_\odot$ yr$^{-1}$ \citep{olivier01} are
much redder, with $(J-K)_{0}$ ranging from 2 to 6.5 mag. The overall
distribution (Fig.\ \ref{fig:jhkLPV.ps}) appears to form a sequence of
ever redder colours with increasing mass-loss rate, a trend that is
well reproduced, e.g., by a model for an M10 type AGB star with
increasing shell opacity \citep{groenewegen93}. Thus, a higher mass
loss has the same effect on $(J-K)_{0}$ and $(H-K)_0$ colours as more
interstellar absorption/reddening, thus making a distinction between
intrinsic and interstellar reddening in these colours impossible.

In contrast, in the $(J-K)_0$ vs. $(K-[12])_0$ and the $(H-K)_0$
vs. $(K-[12])_0$ diagrams (Figs.\ \ref{fig:jk12LPV.ps} and
\ref{fig:hk12LPV.ps}), a separation between interstellar and
circumstellar extinction could theoretically be made. However, this is
hampered by the large dispersion of the Mira stars around the
colour-colour fiducial sequence (0.5 mag), which is due to the
non-contemporaneity of the near- and mid-infrared data and to the
dependence of such relations on metallicity and stellar spectral
type. The uncertainty of interstellar extinction estimates found by
shifting a photometry point along the reddening vector onto the
$H-K,K-[12]$ curve is larger than \Ak $= 1$ mag for $K-[12]>3.5$ mag.

In the $(H-K)_0$ vs. $(K-[12])_0$ plane known Mira stars appear to lie
below the fiducial colour-colour sequence of IRAS-selected oxygen-rich
AGB stars \citep{fouque92,guglielmo93} derived by \citet{vanloon98}.
The offset could be due to the $H$ band water absorption bands
\citep{frogel87,glass95} that are found to be strong in LAV
stars. However, we did not find a correlation between colour and
variability index for these IRAS-selected stars.  The offset might
also suggest a problem transforming the original ESO photometry of
such cold stars in the SAAO system.

Mira stars with $0.2< (H-K)_0 < 3$ mag fit
$$(K-[12])_0 = 4.26(\pm 0.04) + 5.95(\pm0.17) {\rm ~log}(H-K)_0,$$
with an rms deviation of 0.5 mag.  Although we use the SAAO system
\citep{carter90}, in Appendix A we show that this fit is also
valid in the 2MASS photometric system.

\section{Interstellar extinction and intrinsic colours of the SiO targets}
\label{siotargets}
\begin{figure*}[th!]
\begin{center}
\resizebox{0.8\hsize}{!}{\includegraphics{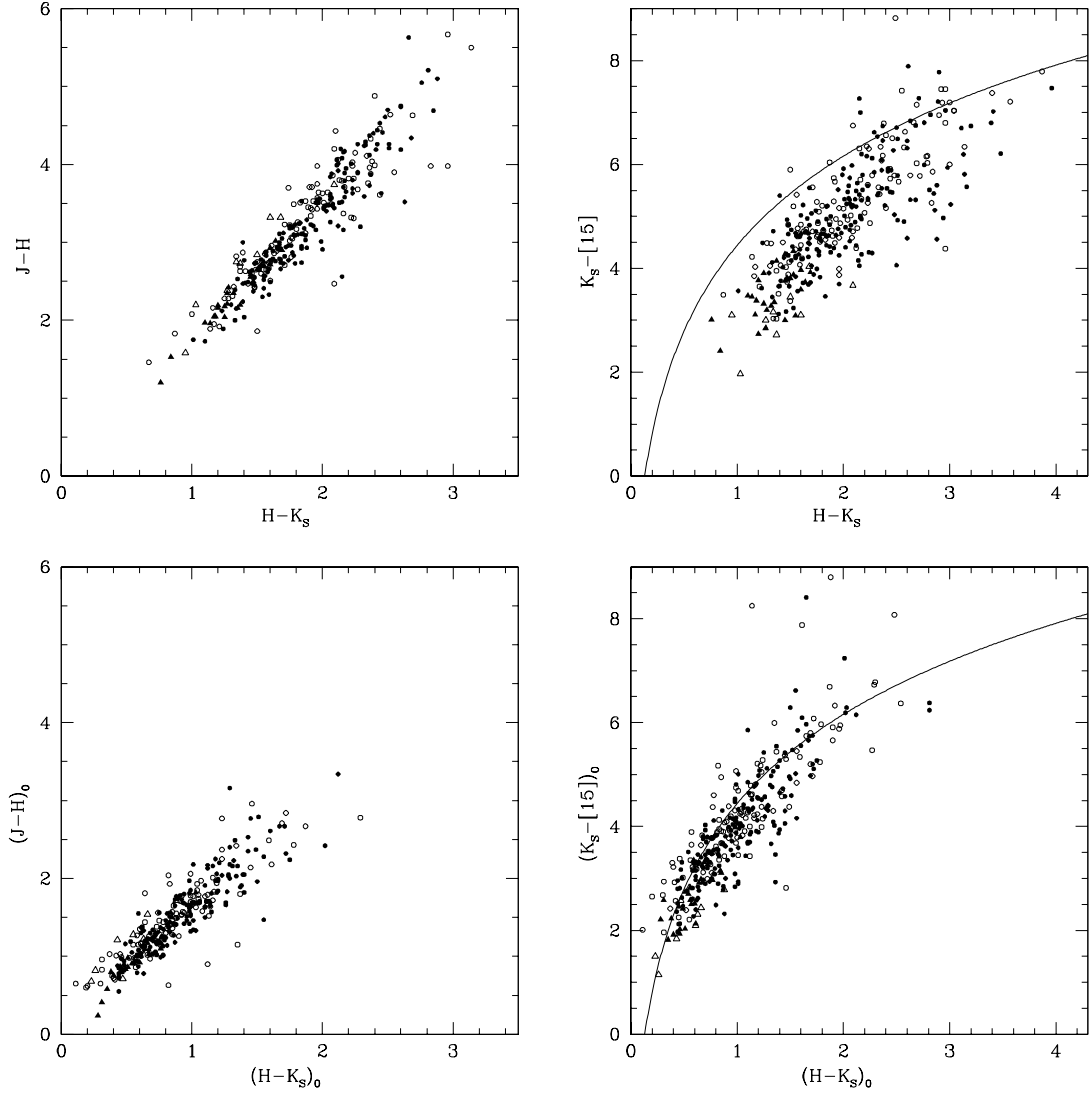}}
\caption{\label{fig:undcol.ps} {\bf Left upper panel:} 2MASS $J-H$
versus $H-$\ks\ colours.  Stars with upper magnitude limits are not
shown.  Dots and triangles represent objects with \ks\ smaller and
larger than 6.0 mag, respectively. Filled and open symbols represent
SiO detections and non-detections, respectively. ``Foreground
objects'' were removed.  The arrow shows the reddening vector for
\Aks\ $= 1$ mag.  {\bf Right upper panel:} 2MASS \ks$-[15]$ versus
$H-$\ks\ colours.  The curve represents the best fit to the colours of
known Mira stars (see Sect.\ \ref{miras}).{\bf Left lower panel:}
Dereddened 2MASS $(J-H)_0$ versus ($H-$\ks$)_0$ colours.  {\bf Right
lower panel:} Dereddened 2MASS (\ks$-[15])_0$ versus ($H-$\ks$)_0$
colours.  }
\end{center}
\end{figure*}

Due to interstellar extinction along the line of sight, for a given
\ks$-[15]$, the $H-$\ks\ colours of the SiO targets are much redder
than those expected from the colour-colour relation of known Mira
stars (upper panels of Fig.\ \ref{fig:undcol.ps}).

The distribution of the SiO targets in the dereddened $(H-K_{\rm
S})_0,(K_{\rm}-[15])_0$ diagram (lower panels of Fig.\
\ref{fig:undcol.ps}) using the ``field'' extinction values, approaches
that of known Mira stars. Therefore, the ``field'' median extinction
is a good approximation of the interstellar extinction for most of the
SiO targets.  Though the dispersion of individual field star
extinctions along a given line of sight is considerable (from 0.1 to
0.8\sig), the distribution is strongly peaked, especially in the Bulge
region. Furthermore, the lifetime of a star on the AGB evolutionary
phase is very short: it is about $5$\% of the time spent on the helium
core burning phase and from 0.1 to 2\% of the time spent on the main
sequence phase of such a star \citep{vassiliadis93}. We therefore
expect most AGB stars to be located in the region with the highest
stellar density along the line of sight.

There is still an asymmetry in the distribution of the SiO targets
around the fiducial colour line of known Miras, which suggests that we
could have underestimated the interstellar extinction for part of the
sample up to 15\%.  In regions of high extinction, Mira-like stars are
detectable at larger distances than ordinary field stars due to their
high near-infrared luminosity.  Deeper infrared observations are
needed to obtain more accurate extinction estimates \citep{figer04}.

\subsection{``Foreground objects''}
\label{foreground}

We can measure the total (circumstellar plus interstellar) extinction
of each SiO target (Table \ref{table:final}), by assuming a stellar
photospheric $(J-K_{\rm s})_0$ colour of 1.4 mag and an $(H-K_{\rm
S})_0$ colour of 0.5 mag (cf. Fig.\ \ref{fig:jhkLPV.ps}).

The difference between these total extinction estimates and 2MASS and
DENIS $J,K_{\rm s}$ data gives an rms of $\Delta$\Aks\ $ = 0.2$ mag,
while estimates from 2MASS $H-K_{\rm s}$ and $J-K_{\rm s}$ colours
give an rms difference of $\Delta$\Ak\ $ = 0.13$ mag. For only 12 SiO
targets we do not have any observed $H-K_{\rm s}$ or $J-K_{\rm s}$
values to determine the total extinction.

As expected, on average the target stars show larger total extinctions
than their surrounding field stars. This is not the case, however, for
a group of $\sim50$ mostly very bright target stars (\ks\ $< 6.0$) at
various longitudes, marked with flag Fg$=1$ in Table
\ref{table:final}, which have total extinctions lower (a least 1\sig)
than the ``field'' extinction and are therefore likely to be
foreground objects.  We dereddened the ``foreground objects'' by
directly shifting them on the $H,K_{\rm s},[15]$ colour-colour
sequence.

\section{Intrinsic colours and mass-loss rates}\label{massloss}
The stellar mass-loss rate is best estimated from measurements of CO
rotational lines.  The CO emission arises in the circumstellar shell.
However, because of confusion with interstellar CO emission, it is
difficult to obtain such measurements toward stars in the inner Galaxy
\citep{winnberg91}.  Although infrared emission also arises in the
stellar photosphere, stellar outflows may be studied using the infrared
emission of dust grains that form in the cool circumstellar
envelopes.  Relations between the infrared colours (e.g.  $J-K$,
$K-L$, $K-[12]$ or $K-[15]$) of O-rich AGB stars and their mass-loss
rate have been established empirically
\citep[e.g.][]{whitelock94,olivier01,alard01} and supported by
theoretical models
\citep[e.g.][]{groenewegen93,ivezic99,jeong03,ojha03}.

The empirical relation between the $(K-[15])_0$ colour and mass loss
rate, $\dot M$, is very useful to study stars detected in the 2MASS or
DENIS surveys and in the ISOGAL or MSX surveys toward the most
obscured regions of the Galaxy.

The uncertainties arising from the variability of the stars and the
temporal difference between the \ks\ and 15$\mu$m measurements are
somewhat alleviated by using an average of the 2MASS and DENIS \ks\
fluxes and of the ISOGAL and MSX 15$\mu$m measurements.  The remaining
r.m.s. uncertainty of the mass-loss rate is thus a factor $\sim$2 for
$\dot M > 10^{-6}$ M$_\odot$ yr$^{-1}$ \citep{ojha03}.

\begin{figure}[th!]
\begin{center}
\resizebox{0.8\hsize}{!}{\includegraphics{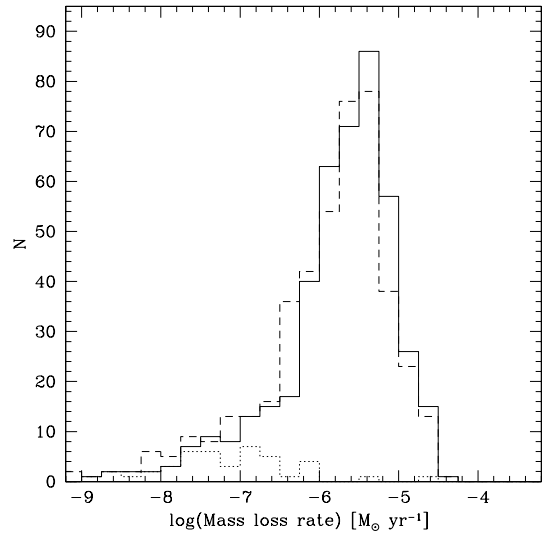}}
\caption{\label{fig:jeong.ps} Distribution of mass-loss rates derived
from the $K-[15]$ vs.\ $\dot M$ relation \citet{jeong03}. The
continuous line shows the distribution for all SiO targets dereddened
using Lutz's extinction law (Curve 3) and the dotted line that of
 foreground stars.  The dashed line is the
distribution of all SiO targets using the mid-infrared extinction law
of Mathis (Curve 1).}
\end{center}
\end{figure}

Following the prescription of \citet{jeong03} and \citet{ojha03} we
obtained mass-loss rates for the SiO targets, the distribution of
which is shown in Fig. \ref{fig:jeong.ps}. 90\% of the sources have
implied mass-loss rates between $10^{-7}$ and $2 \times 10^{-5}$
M$_\odot$ yr$^{-1}$, with a peak in the range $10^{-6}$ -- $10^{-5}$
M$_\odot$ yr$^{-1}$, although the apparent distribution is widened by
the uncertainty of the mass-loss--colour relation, by the photometric
uncertainty, and by the effects of variability.

Our selection criterion on (\ks$-[15])_0$ for the ISOGAL sample
completely eliminated sources with $\dot M >10^{-5}$, while the
elimination of OH/IR sources and the criteria on $A-D$ and $C-E$
colours for the MSX sample also considerably reduced the proportion of
sources with large mass-loss rates
\citepalias[see][]{messineo03_2,messineo02}.  The same results are
obtained considering the subsample of targets with the best extinction
corrections, i.e.\ with $\sigma _{\rm A_{\rm K_{\rm s}}}<0.2$
mag. Furthermore, the distribution of the mass-loss rates of the
targets with detected SiO maser emission appears similar to those
without detected SiO \citepalias{messineo02}.

A possible underestimation of the interstellar extinction results in
an overestimation of the mass-loss rate. Adopting Mathis' mid-infrared
extinction law rather than Luts's, the distribution of the mass-loss
rates only slightly shifts toward lower values (see Fig.\
\ref{fig:jeong.ps}).

\section{Conclusion}\label{conclusion}
 We estimated the interstellar extinction toward 441 SiO maser AGB
 stars. From the 2MASS stars within 2-4\arcmin\ of each SiO star, we
 estimated a mean extinction toward the SiO stars.  We therefore
 shifted the colours of field stars $(J-K_{\rm s})$ and $(H-K_{\rm
 s})$ versus $K_{\rm s}$ magnitudes along the reddening vector onto
 the reference RGB. The use of both colour-magnitude planes enabled us
 to obtain a mean extinction for each field and new constraints on the
 index of the near-infrared extinction power law $\alpha$.  We found
 that a value of $\alpha = 1.6$ is inconsistent with the colours of
 inner Galactic stars, and by taking 47 Tuc as a reference for the RGB
 we find $\alpha = 1.9 \pm 0.1$.
  
 For \ks-band extinctions larger than 1.6 mag the 2MASS $(K_{\rm
 S},J-K_{\rm s})$ CMD provides extinction estimates that are too low
 due to a selection effect from the $J$-band dropout of more distant
 sources.  The 2MASS ($K_{\rm s},H-K_{\rm s}$) CMD suffers less from
 this bias.

 We reviewed near- and mid-infrared dereddened colour-colour relations
 of Mira stars and use them to test the quality of the extinction
 corrections for each SiO target.  Under the assumption that the SiO
 target stars are spatially distributed similar to the surrounding
 field stars, we corrected the photometric measurements of the SiO
 targets by adopting the median extinction of their surrounding field
 stars.  The derredened colours of the SiO targets are not
 symmetrically distributed around the fiducial colour-colour line of
 known Mira stars, which suggests that for part of the SiO targets we
 may still be underestimating the interstellar extinction by up to
 15\%. About 50 SiO targets lie significantly in the ``foreground'' of
 the mean stellar distribution.
 
 Using the relation between mass-loss rate and (\ks$-15)_0$ colour
 given by \citet{jeong03}, we estimated that most of the SiO targets
 have mass-loss rates in the range 10$^{-7}$ to 10$^{-5}$ M$_{\odot}$
 yr$^{-1}$.

\begin{acknowledgements}
The MSX transmission curves were kindly provided by M. Egan.  MM
thanks P. Popowski, J. van Loon, S. Ganesh, and M. Schultheis for
useful discussions about interstellar extinction, and M. Sevenster for
her constructive criticism.  This paper uses and partly depends on the
studies of Mira stars conducted at the SAAO observatory by P.
Whitelock and her collaborators.  This publication makes use of data
products from the IRAS data base server, from the Two Micron All Sky
Survey, from the DEep Near-Infrared Survey of the southern sky, from
the Midcourse Space Experiment, and from the SIMBAD data base.  The
work of MM was funded by the Netherlands Research School for Astronomy
(NOVA) through a {\it netwerk 2, Ph.D. stipend}.
\end{acknowledgements}

\appendix
\section{SAAO and 2MASS colours and magnitudes}
Transformation equations between the colours and magnitudes measured
in the SAAO \citet{carter90} and 2MASS photometric systems have been
derived by \citet{carpenter01} using a list of mostly blue 94
photometric standards. Figure 12 in \citet{carpenter01} shows that the
differences between  magnitudes and colours obtained with the two
systems are smaller than 0.15 mag.

Considering that Mira stars typically have a pulsation amplitude in
the near-infrared of 1-2 mag and that 2MASS data are from a
single-epoch observation randomly taken with respect to the stellar
phase, the system transformations only have a secondary effect on the
total colour and magnitude uncertainty, when comparing data from 2MASS
with data taken with the SAAO telescope. However, since Mira stars are
cold objects, we must exclude the idea that a combination of molecular
bands and filter transmissions could generate a different colour
transformation for this special class of objects.  We therefore looked
for 2MASS counterparts of the 104 outer Bulge Mira stars monitored by
\citet{whitelock94}.  Mira stars are among the brightest objects
detected in the $K_{\rm s}$ band such that identification of their
2MASS counterparts is straightforward \citepalias{messineo03_2} (only
three sources were excluded because they had no unique counterparts).

Differences between the mean magnitudes obtained with SAAO
observations \citep{whitelock91} and the single-epoch 2MASS data have
a dispersion of up to 0.8 mag and the following mean differences:
$$ {K_{\rm s}}_{(\rm 2MASS)} - K_{(\rm SAAO)} =  -0.15 \pm 0.06 {\rm ~mag};$$
$$(J-K_{\rm s})_{(\rm 2MASS)} - (J-K)_{(\rm SAAO)} = -0.14 \pm 0.05 {\rm ~mag};$$
$$ (H-K_{\rm s})_{(\rm 2MASS)} - (H-K)_{(\rm SAAO)} = -0.06 \pm 0.03 {\rm ~mag}.$$

The colour-colour relations  described in Sect.\
\ref{miras} using SAAO photometry also  describe the 2MASS
sequence of Mira stars well.

\section{IRAS and MSX filters}
Most of the past work was carried out using IRAS photometry;
therefore, the currently available colour-colour relations of Mira
stars use mid-infrared data from the IRAS catalogue.  A comparison of
mid-infrared filters is therefore mandatory when translating old
findings into new MSX and ISOGAL colours.

\begin{figure}[ht!]
\begin{centering}
\resizebox{0.8\hsize}{!}{\includegraphics{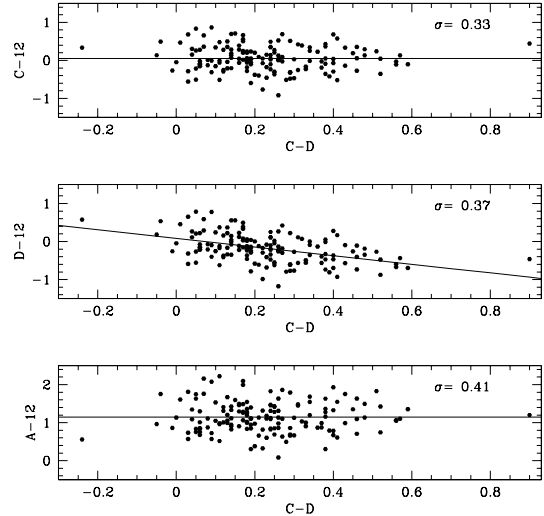}}
\caption{\label{fig:cc2.ps} Colour-Colour diagrams of our SiO targets.
The IRAS 12$\mu$m magnitude is defined as $[12]=-2.5 \log F_{12}[{\rm
Jy}]/28.3$. The continuous lines are our best fits.}
\end{centering}
\end{figure}

Figure \ref{fig:cc2.ps} shows the difference between MSX magnitudes
and IRAS 12 $\mu$m magnitude for the SiO targets.  Note that the $D$
filter excludes the silicate feature around 9.7 $\mu$m, while $A$ and
$C$ filters include part of it (see Fig. \ref{fig:filter.ps}).
Therefore the ($D-[12])$ colour shows a dependence on the ($C-D$)
colour, which increases when the silicate feature at 9.7 $\mu$m starts
to be self-absorbed.  The ($A-[12]$) and ($C-[12]$) colours do not
show any trend with the ($C-D$) colour.  Due to both uncertainties of
photometric measurements and source variability, scatter is large;
however we derived relations between the $A$, $C$, and $D$ and the
[12] magnitudes, as follow:\\ $A-[12]=1.15\pm0.03$ mag\\
$C-[12]=0.05\pm0.03$ mag\\$D-[12]=0.08(\pm0.18) -1.13(\pm0.05)(C-D)$ mag.

\end{document}